\documentclass[aps,prl,showpacs,showkeys,superscriptaddress,nofootinbib,a4paper,twocolumn]{revtex4}

\usepackage{amsfonts, epsf}

\begin{document} 

\title{Non-constant ground states for symmetry-breaking fields\\in brane world models}

\author{Alan Knapman}
\email[Email: ]{a.h.knapman@ncl.ac.uk}
\author{David J. Toms}
\email[Email: ]{d.j.toms@ncl.ac.uk}
\affiliation{School of Mathematics and Statistics, University of Newcastle Upon Tyne, Newcastle Upon Tyne NE1 7RU, England}

\begin{abstract}
We obtain an approximate analytical solution for the ground state of a bulk scalar field with a double-well potential in the Randall-Sundrum brane world background, in a situation where the boundary conditions rule out a constant field configuration except for the zero solution. The stability of the zero solution is determined by the brane separation. We find our approximation near the critical separation at which the zero solution becomes unstable to small perturbations.
\end{abstract}


\maketitle

\today

\newcommand{\nn}{\nonumber\\}
\newcommand{\be}{\begin{equation}}
\newcommand{\ee}{\end{equation}}
\newcommand{\bea}{\begin{eqnarray}}
\newcommand{\eea}{\end{eqnarray}}

\vspace{5mm}

\noindent
The idea that our universe might be a membrane embedded in some higher dimensional spacetime has been receiving a lot attention in recent years. In particular, Randall and Sundrum \cite{Randall:1999ee} have proposed a 5-dimensional model where the extra dimension has an orbifold compactification, with two flat 3-branes with opposite tensions sitting at two orbifold fixed points.

In the original Randall-Sundrum model, all fields except for gravity are confined to the branes. However, the possibility of other fields in the ``bulk'' space between the branes has been considered by many authors, in the context of: stabilisation of the distance between the branes, e.g.~\cite{stability}; self-consistency, e.g.~\cite{kanti, knapman, flachitoms}; cosmology, e.g.~\cite{cosm} and models in which all fields propagate in the bulk spacetime, e.g.~``universal extra dimensions'' \cite{Appelquist:2000nn}.

In the last example, it is reasonable to ask what happens when a field that may undergo spontaneous symmetry-breaking, such as a Higgs field, exists inside the bulk. In the standard model, this is what provides particles with their masses. Symmetry-breaking in brane world models has also been investigated in e.g.~\cite{symmbr}. A simple model for such a field is a real scalar field $\Phi$ with the double-well potential

\be
  U(\Phi) = \frac{\lambda}{24}(\Phi^2-a^2)^2. \label{2.2}
\ee

This potential has a maximum at $\Phi=0$ and minima at non-zero values of $\Phi=\pm a$. It is straightforward to show that these constant values represent possible ground states of the field. In ordinary Minkowski spacetime, which is homogeneous and isotropic, if the zero solution is unstable to small perturbations due to the dynamics of the model, then the stable ground states are indeed these non-zero constant values.

It is possible to construct homogeneous and isotropic spacetimes in which the ground state may not be a constant. The example of a toroidal spacetime was provided by Avis and Isham \cite{avis}, who showed that if the torus is below a critical size set by the parameters of the potential, the zero solution is the stable ground state; above that critical size, the stable state is a spatially dependent solution.

Similarly, as we illustrate in this Letter, the boundary conditions in the Randall-Sundrum model can rule out constant non-zero field values. Again, there is a critical parameter, in this case the size of the extra dimension, that determines the stability of the zero ground state and a state of broken symmetry must be found. This symmetry-broken ground state will necessarily be non-constant. This could then potentially provide some very interesting particle physics in the next round of collider experiments, as particle ``masses'' would become a function of the extra-dimensional coordinate.

In this Letter, we use a method presented in \cite{david} for obtaining approximate analytical expressions for ground states in situations where the boundary conditions rule out constant field configurations. We consider a simple yet interesting toy model in which a single scalar field exists in the background of the Randall-Sundrum spacetime. For generality, we have $D$ brane dimensions and one extra dimension. The field is in the so-called ``twisted'' configuration (see below), which is especially simple. This configuration is equivalent to imposing Dirichlet (vanishing-field) boundary conditions on the branes.

The $(D+1)$ dimensional Randall-Sundrum metric with signature $(+-...\,-)$ is
\be
  ds^2 = \hat{g}_{\hat{\mu} \hat{\nu}} d\hat{x}^{\hat{\mu}} dx^{\hat{\nu}} = e^{-2k|y|} \eta _{\mu \nu} dx^\mu dx^\nu - dy^2 , \label{1.1}
\ee
where a caret denotes a $(D+1)$ dimensional quantity. A non-zero constant $k$ gives the ``warping'' of the bulk space, so that the higher dimensional space is a slice of AdS$_{D+1}$ spacetime. The extra dimension $y$ lies between the branes at $y=0$ (the ``hidden'' brane) and $y=\pm\pi r$ (the ``visible'' brane, which represents our universe). A $\mathbb{Z}_2$ symmetry is imposed identifying points at $y$ and $-y$. All of the other spacetime directions, labelled by coordinates $x^\mu$, are infinite and flat.

The action $S$ of a real bulk scalar field $\Phi$ is chosen to be
\be
  S[\Phi] = \int dt\int d^{D-1} x\int dy\,\sqrt{\hat{g}} \left[ \frac{1}{2}\hat{g}^{\hat{\mu } \hat{\nu }} \partial _{\hat{\mu}} \Phi \partial _{\hat{\nu}} \Phi - U(\Phi) \right]. \label{2.1}
\ee
where $\hat{g} = |\!\det \hat{g}_{\mu\nu}|$. $U(\Phi)$ is the double-well potential in Eq.~(\ref{2.2}).

The field equation for the ground state $\phi$, which depends only on $y$, following from Eq.~(\ref{2.1}) is
\be
  \Delta\phi - U'(\phi) = 0 \label{2.6}
\ee
where
\be
  \Delta = e^{D\sigma}\frac{d}{dy}\left( e^{-D\sigma}\frac{d}{dy}\right) . \label{2.8}
\ee

As is pointed out in \cite{flachitoms}, the orbifold nature of the extra dimension means that the field can have two kinds of configuration: ``twisted'', $\phi(y)=-\phi(-y)$ and ``untwisted'', $\phi(y)=+\phi(-y)$.

The twisted field configuration, together with the requirement that the field is continuous across the branes, is equivalent to imposing Dirichlet boundary conditions on the branes, i.e.~$\phi(0)=\phi(\pi r)=0$.

For the potential (\ref{2.2}), $\phi=0$ and $\phi=\pm a$ are solutions to Eq.~(\ref{2.6}). There are two key issues that then arise. The first is whether or not the boundary conditions are satisfied. We note that our Dirichlet boundary conditions prohibit a constant field, except if the field is everywhere zero. Therefore, the standard Minkowski spacetime ground states $\phi=\pm a$ are not allowed. The second is that the solution must be stable to perturbations.

Following the method described in \cite{david}, which involves looking at the energy of a perturbed solution $\phi+h$, with $h$ treated as small, we find a critical value of the brane separation $r=r_c$ at which $\phi=0$ becomes unstable, given by
\be
  r_c = \left[ \frac{\lambda a^2}{6} - \frac{D^2k^2}{4} \right]^{-1/2}. \label{2.16}
\ee
The field $\phi=0$ is unstable if $r>r_c$. The stable ground state is then necessarily one of non-constant field configuration.

In \cite{LauraToms}, a systematic method was described for obtaining approximate solutions for the ground state in cases where it is not possible to solve the equation of motion exactly. This method was adapted to cavities in \cite{david}.

We take the size of the extra dimension $r$ to be slightly greater than the critical value:
\be
  r=(1+\epsilon)r_c\;,\label{2.17}
\ee
with $0<\epsilon\ll 1$. We also transform to dimensionless coordinates $z = y/\pi r$. The branes are then at $z=0$ and $z=1$. The operator $\Delta$ can be expanded in powers of $\epsilon$ to give
\be
  \Delta=\Delta_c+\epsilon\Delta_1+\epsilon^2\Delta_2+\cdots \;,\label{2.18}
\ee
where $\Delta_c$ denotes the original operator evaluated with $r=r_c$. The operators $\Delta_1,\Delta_2,\ldots$ are straightforward to obtain. The first is given by
\be
  \Delta_1 = \frac{Dk}{y_c}\frac{d}{dz} - \frac{2}{y_c^2}\frac{d^2}{dz^2}, \label{2.19}
\ee

\begin{figure}[htb]
\begin{center}
\leavevmode
\epsfxsize=70mm
\epsfysize=50mm
\epsffile{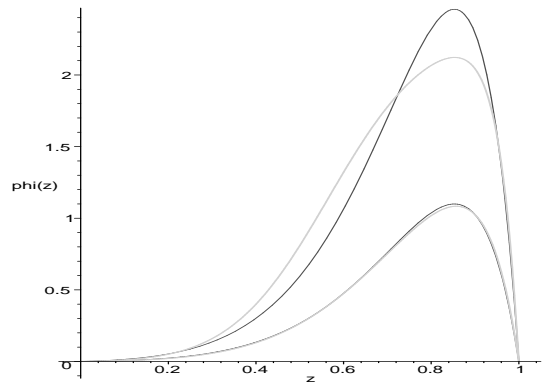}
\end{center}
\caption{\footnotesize Comparison of the approximate solutions (black curves) with the numerical solutions (grey curves). The lower pair of curves, for $\epsilon=0.1$, illustrates the excellent agreement between the two. The upper pair of curves, for $\epsilon=0.5$, shows that the approximation is beginning to break down for this value of $\epsilon$. The parameters chosen are $D=4$, $k=1$, $\lambda=5.2$ and $a=2.4$.
}\label{fig1}
\end{figure}

It can be shown (see \cite{LauraToms}) that the field $\phi$ can be expanded as
\be
  \phi(z) = \epsilon^{1/2}\left\lbrack\phi_0(z) +\epsilon\phi_1(z)+\epsilon^2\phi_2(z) +\cdots\right\rbrack\;,\label{2.20}
\ee
where $\phi_0(z),\phi_1(z),\ldots$ are independent of $\epsilon$. Substituting this into the equation of motion (\ref{2.6}) and equating coefficients of equal powers in $\epsilon$, we obtain a set of coupled differential equations. The first of the set,
\be
  \Delta_c\phi_0+\frac{\lambda a^2}{6}\phi_0 = 0\;,\label{2.21}
\ee
gives the first order approximation $\phi_0$ to the non-constant ground state $\phi$. 

The overall scale is left undetermined, as the equation is homogeneous. If we let $\tilde{\phi}_0(z)$ be any solution of Eq.~(\ref{2.21}) that satisfies $\tilde{\phi}_0(0)=\tilde{\phi}_0(1)=0$, then we can write $\phi_0(z) = A\tilde{\phi}_0(z)$, for some constant $A$. The overall scale $A$ is determined by requiring the field expanded as in Eq.~(\ref{2.20}) to minimise the energy to lowest order in $\epsilon$, since we are dealing with the ground state. After some calculation, this gives
\bea
  A = \sqrt{6\int_0^1dz\,e^{-\pi Dkr_cz}\,\tilde{\phi}_0\Delta_1\tilde{\phi}_0\over\lambda\int_0^1dz\,e^{-\pi Dkr_cz}\,\tilde{\phi}_0^4}\;.\label{2.27}
\eea
It is simple to find a solution to Eq.~(\ref{2.21}) that satisfies the boundary conditions. After calculating the overall scale, we find
\bea
  \phi(z) &\approx& \epsilon^{1/2} \left[ \frac{\pi Dk\left(D^2k^2r_c^2+4\right)\left(D^2k^2r_c^2+16\right)}{4\lambda r_c\left(e^{\pi Dkr_c}-1 \right)} \right] ^{1/2} \nn
           && \times\;e^{\pi Dkr_cz/2}\sin(\pi z)\; \label{2.30}
\eea
for the first order approximation to the ground state. We have also calculated the next order correction, though the result is too lengthy to report here.

To both check the approximation and to investigate the behaviour of the solution for regions in which the approximation is not valid (i.e. values of $\epsilon>1$), we have plotted the numerical solution to the field equation (\ref{2.6}) for a certain set of parameters and with various values of $\epsilon$. The numerical solution was obtained using a shooting method procedure.

Figure \ref{fig1} illustrates the good agreement between the approximate and numerical solutions for the small value of $\epsilon$ of $0.1$. This agreement starts to break down around $\epsilon=0.5$, as also shown in Figure \ref{fig1}. Considering that the approximation is only given to the lowest order, the degree of accuracy over such a range is quite remarkable. It is possible to improve the agreement between the numerical result and the analytical approximation by including the next order correction to (\ref{2.30}).

\begin{figure}[htb]
\begin{center}
\leavevmode
\epsfxsize=70mm
\epsfysize=50mm
\epsffile{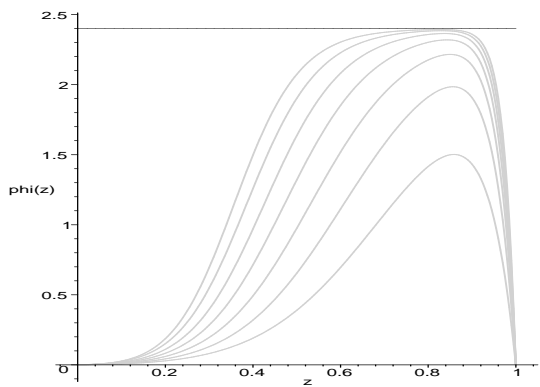}
\end{center}
\caption{\footnotesize Behaviour of the numerical solution for a range of values of $\epsilon$ from $0.2$ up to $1.4$ in steps of $0.2$. The smallest curve corresponds to $\epsilon=0.2$, with each progressively larger curve representing a step upwards in $\epsilon$. The horizontal black line is the constant solution $\phi=a$, which the curves are tending towards away from the branes.}\label{fig3}
\end{figure}

In Figure \ref{fig3}, we present the behaviour for a range of values of $\epsilon$ from $0.2$ up to $1.4$. This illustrates the interesting behaviour that, as $\epsilon$ gets larger, the ground state gets closer to resembling the constant $\phi=a$ solution. The function is brought down to zero at each brane as dictated by the boundary conditions. It seems reasonable to assume that the solution tends towards a step function in the limit that $\epsilon\rightarrow\infty$.

{\em Conclusions.}---In this Letter, we have obtained an approximate analytical solution for the ground state of a scalar field with a double-well potential existing in the bulk of the Randall-Sundrum spacetime. The field is in the so-called ``twisted'' configuration, which is the simplest situation in which the boundary conditions rule out a constant field configuration except for the zero solution. The stability of the zero solution to small perturbations is determined by the brane separation. We find a critical separation at which the zero solution becomes unstable, which is given by Eq.~(\ref{2.16}).

Our approximation, given in Eq.~(\ref{2.30}), applies in the regime in which the brane separation exceeds the critical value by a small amount. In fact, numerical results show that approximation works up to separations that are greater than the critical value by nearly 50\%.

The general form of these near-critical solutions is of a sinusoid multiplied by an exponential. For larger brane separations, the solutions begin to distort from this simple shape and begin to tend towards a step-like function, which is as close as the boundary conditions will allow to the constant $\phi=a$ solution that is the more familiar case. Of course, at or below the critical separation, the ground state is just $\phi=0$.

The most interesting behaviour is therefore to be found exactly in the region in which the approximation applies. In this region, the deviation from constant behaviour is the greatest.

If the field is interpreted as a simple model of a Higgs field that will generate particle masses in the bulk, such as in models of universal extra dimensions, this would cause these ``masses'' to vary greatly with position along the extra dimension. This would certainly give rise to some interesting particle physics that may be measurable.

In the simple example given here, such a field will disappear on the branes, indicating that a model with universal extra dimensions cannot use twisted conditions, unless extra boundary terms are included. It would therefore be interesting to investigate these effects in more realistic models of universal extra dimensions and what kinds of particle phenomena might result. The power of the approximation method used here is that it would allow this phenomenology to be analysed much more easily than by dealing with the exact solution to the field equation.

{\em Acknowledgement.}---We would like to thank Laila Alabidi for checking some results.


\begin{thebibliography}{99}

\bibitem{Randall:1999ee}
L.~Randall and R.~Sundrum,
%``A large mass hierarchy from a small extra dimension,''
Phys.\ Rev.\ Lett.\  {\bf 83} (1999) 3370.

\bibitem{stability}
W.~D.~Goldberger and M.~B.~Wise, Phys.~Rev.~Lett.~{\bf 83} (1999) 4922;
J.~Garriga, O.~Pujol\`{a}s, and T.~Tanaka, Nucl.~Phys.~B {\bf 605} (2001) 192;
W.~D.~Goldberger and I.~Z.~Rothstein, Phys.~Lett.~B {\bf 491} (2000) 339.

\bibitem{kanti}
R.~Hofmann, P.~Kanti and M.~Pospelov, Phys.~Rev.~D {\bf 63} (2001) 124020.

\bibitem{flachitoms}
A.~Flachi and D.~J.~Toms,
%``Quantized bulk scalar fields in the Randall-Sundrum brane-model,''
Nucl.\ Phys.\ B {\bf 610} (2001) 144.

\bibitem{knapman}
A.~Knapman and D.~J.~Toms,
%``Stress-energy tensor for a quantised bulk scalar field in the Randall-Sundrum
%brane model,''
Phys.\ Rev.\ D {\bf 69} (2004) 044023.

\bibitem{cosm}
P.~Binetruy, C.~Deffayet, U.~Ellwanger and D.~Langlois,
%``Brane cosmological evolution in a bulk with cosmological constant,''
Phys.\ Lett.\ B {\bf 477} (2000) 285;
S.~Mukohyama,
%``Brane-world solutions, standard cosmology, and dark radiation,''
Phys.\ Lett.\ B {\bf 473} (2000) 241;
D.~Ida,
%``Brane-world cosmology,''
JHEP {\bf 0009} (2000) 014;
Y.~Himemoto and M.~Sasaki,
%``Brane-world inflation without inflaton on the brane,''
Phys.\ Rev.\ D {\bf 63} (2001) 044015;
S.~Kobayashi, K.~Koyama and J.~Soda,
%``Quantum fluctuations of bulk inflaton in inflationary brane world,''
Phys.\ Lett.\ B {\bf 501} (2001) 157.

\bibitem{Appelquist:2000nn}
T.~Appelquist, H.~C.~Cheng and B.~A.~Dobrescu,
%``Bounds on universal extra dimensions,''
Phys.\ Rev.\ D {\bf 64} (2001) 035002.

\bibitem{symmbr}
A.~Flachi and D.~J.~Toms,
%``Higgs mechanism in the Randall-Sundrum model,''
Phys.\ Lett.\ B {\bf 491} (2000) 157;
T.~Gherghetta and A.~Pomarol,
%``Bulk fields and supersymmetry in a slice of AdS,''
Nucl.\ Phys.\ B {\bf 586} (2000) 141;
S.~J.~Huber and Q.~Shafi,
%``Higgs mechanism and bulk gauge boson masses in the Randall-Sundrum  model,''
Phys.\ Rev.\ D {\bf 63} (2001) 045010.

\bibitem{avis}
S.~J.~Avis and C.~J.~Isham, Proc.~Roy.~Soc.~(London) A, {\bf 363} (1978) 581.

\bibitem{david}
D.~J.~Toms,
%``Ground State Solutions For A Self-Interacting Scalar Field Confined In A Cavity,''
J.\ Phys.\ A {\bf 36} (2003) 5121.

\bibitem{LauraToms}
R.~Laura and D.~J.~Toms,
%``Twisted Goldstone Models,''
J.\ Phys.\ A {\bf 15} (1982) 3725.

\end{thebibliography}
\end{document}